\documentclass[twocolumn,3p]{elsarticle}
\usepackage{subfigure}

\usepackage{natbib}
\usepackage{captcont}
\usepackage[final,		
	colorlinks,			
	linktocpage,		
	linkcolor=blue,		
	citecolor=blue,		
	urlcolor=blue,		
	breaklinks=true,	
	]{hyperref}
\usepackage[all]{hypcap}	

\begin{document}

\begin{frontmatter}

\title{Measurement of camera image sensor depletion thickness with cosmic rays}

\author[WIPAC]{J. Vandenbroucke\corref{corauths}}
\author[Rochester]{S. BenZvi}
\author[WIPAC]{S. Bravo}
\author{K. Jensen}
\author[WIPAC]{P. Karn}
\author[WIPAC]{M. Meehan}
\author[Sensorcast]{J. Peacock}
\author[WIPAC]{M.~Plewa}
\author[Physics]{T. Ruggles}
\author[Columbia]{M. Santander}
\author[WIPAC]{D. Schultz}
\author[Loyola]{A. L. Simons}
\author[WIPAC]{D. Tosi}

\address[WIPAC]{Physics Department and Wisconsin IceCube Particle Astrophysics Center, University of Wisconsin, Madison, WI 53706, USA}
\address[Physics]{Physics Department, University of Wisconsin, Madison, WI 53706, USA}
\address[Rochester]{Department of Physics and Astronomy, University of Rochester, Rochester, NY, USA}
\address[Columbia]{Department of Physics and Astronomy, Barnard College, Columbia University, New York, NY, USA}
\address[Sensorcast]{Sensorcast}
\address[Loyola]{Seaver College of Science and Engineering, Loyola Marymount University, Los Angeles, CA 90045, USA}

\begin{abstract}
Camera image sensors can be used to detect ionizing radiation in addition to optical photons.  In particular, cosmic-ray muons are detected as long, straight tracks passing through multiple pixels.  The distribution of track lengths can be related to the thickness of the active (depleted) region of the camera image sensor through the known angular distribution of muons at sea level.  We use a sample of cosmic-ray muon tracks recorded by the Distributed Electronic Cosmic-ray Observatory to measure the thickness of the depletion region of the camera image sensor in a commercial smart phone, the HTC Wildfire S.  The track length distribution prefers a cosmic-ray muon angular distribution over an isotropic distribution.  Allowing either distribution, we measure the depletion thickness to be between 13.9~$\mu$m and 27.7~$\mu$m.  The same method can be applied to additional models of image sensor.  Once measured, the thickness can be used to convert track length to incident polar angle on a per-event basis.  Combined with a determination of the incident azimuthal angle directly from the track orientation in the sensor plane, this enables direction reconstruction of individual cosmic-ray events.
\end{abstract}


\cortext[corauths]{Corresponding author: justin.vandenbroucke@wisc.edu}











\end{frontmatter}



\section{Introduction}

The Distributed Electronic Cosmic-ray Observatory (DECO)~\cite{decoWeb} is a network of mobile devices in which camera image sensors are used to detect ionizing radiation including cosmic rays and ambient radioactivity.  Although designed to detect optical photons, cell phone camera image sensors (predominantly using CMOS technology) are also sensitive to ionizing radiation incident on the depletion region.  Consumer technology can therefore be used for purposes similar to those of custom-built trackers used for particle physics and astro-particle physics.  For an overview of DECO, see~\cite{decoICRC}.

Although the area of each sensor is small (typically $\sim$0.15~cm$^2$, varying from model to model), many sensors can be harnessed together.  The area of silicon installed in camera image sensors of cell phones and other consumer devices throughout the world\footnote{Worldwide there are $\sim$10$^9$ cell phones with cameras, each with a sensor area $\sim$10$^1$~mm$^2$.} is at least $\sim$10$^4$~m$^2$.  This is two orders of magnitude larger than the largest professional silicon trackers (installed in the Compact Muon Solenoid~\cite{CMS} and the Fermi Large Area Telescope~\cite{Atwood}).  Each image sensor provides megapixels of resolution for track direction determination and particle identification.  Finally, we benefit from many years of experience of astronomers in identifying, classifying, and removing cosmic-ray tracks from CCD images~\cite{Groom}.  CCDs and CMOS sensors are established detectors of cosmic rays and background radioactivity (which are usually a background to optical or X-ray photon detection) and can potentially detect dark matter as well~\cite{DAMIC2012,DAMICICRC}.


The DECO mobile app\footnote{Available at http://wipac.wisc.edu/deco} records camera images continuously and applies an online filter, implemented in the phone CPU, to select particle candidates.  These data, along with associated metadata, are automatically synchronized to a central server for analysis by scientists and members of the public.  DECO features a web-based data browser\footnote{Available at http://wipac.wisc.edu/deco/data} where users can query the data including quantities such as device ID, model ID, timestamp, and geolocation.  Users can click individual events to view the event image or the event location in Google Maps.  For privacy, latitude and longitude are degraded to 0.01$^\circ$ resolution and image data are provided only in a zoomed, cropped, and false colored version.

There are several challenges to identifying and calibrating images obtained from a heterogeneous network of consumer devices.  Each model, and potentially each device of an individual model, has different noise characteristics.  Furthermore, the noise varies with environmental conditions such as temperature.  Cosmic rays must be discriminated from background events including those due to sensor artifacts, thermal noise fluctuations, and low-energy particle events induced by radioactivity within or near the device.  Finally, cameras in advanced mobile devices include post-processing hardware, firmware, and software designed to remove noise, which can also remove particle events.  There is evidence in the DECO data that particular advanced models have low rats of particle detection because of this noise removal feature.

Using the DECO dataset, we demonstrate discrimination between GeV cosmic-ray muon tracks and MeV electron tracks caused by radioactive decay using topological cuts based on track images.  Using the muon event sample, we fit the track length distribution with a model based on the known cosmic-ray muon angular distribution at sea level and with a single degree of freedom corresponding to the thickness of the depletion region of the camera image sensor.  We also fit the distribution with an analogous functional form corresponding to an isotropic distribution.  The results simultaneously validate the use of cell phone camera image sensors as cosmic-ray muon detectors and provide a measurement of a parameter of camera image sensor performance which is not otherwise publicly available.  We focus on a particular model, the HTC Wildfire S.

\begin{figure}[htbp]
\includegraphics[width=0.5\textwidth]{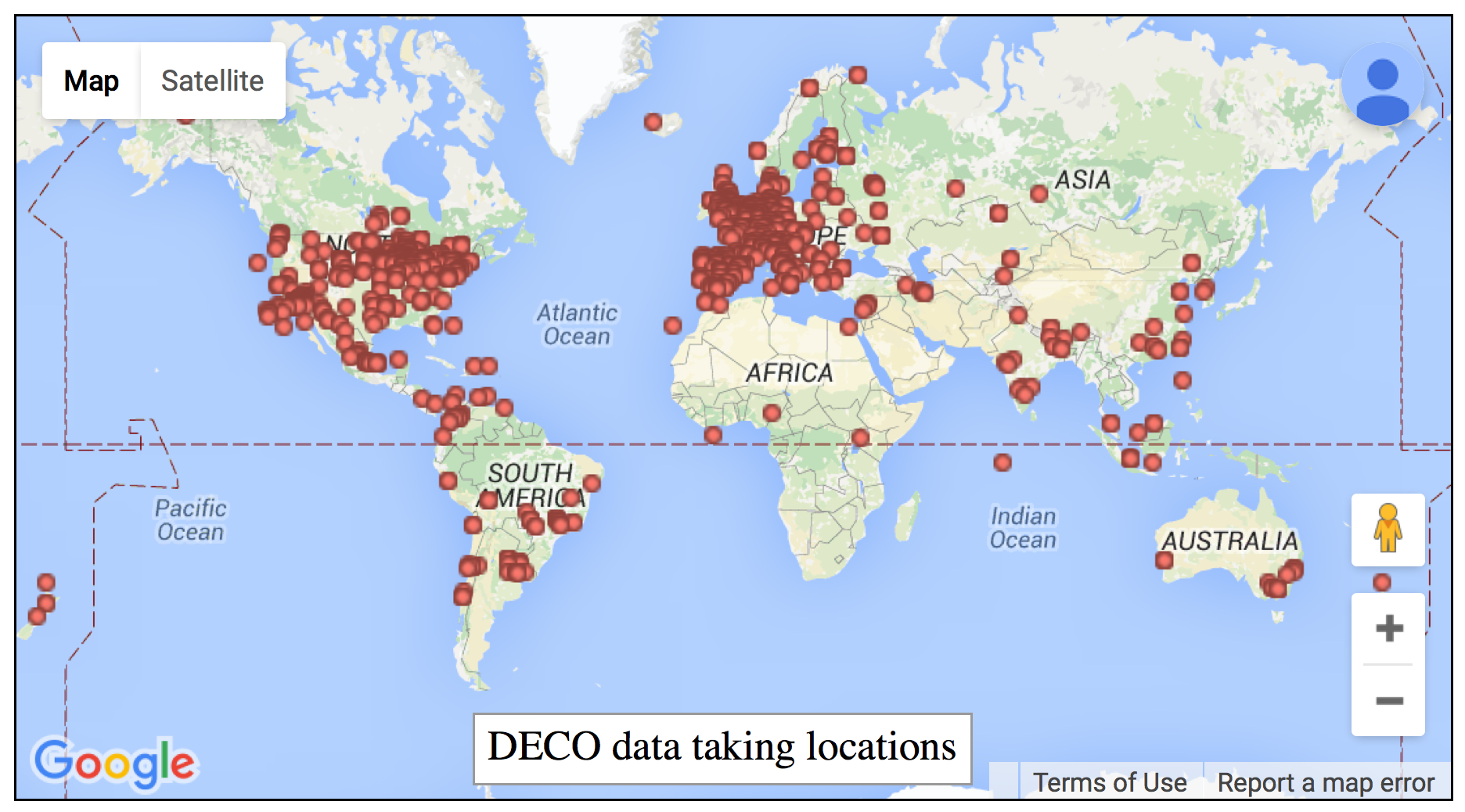}
\caption{Map of DECO data collection locations.  Data have been collected on all seven continents.  \label{map}}
\end{figure}

\section{Data collection}

DECO has been collecting data since September 2011, with the app released publicly in September 2014.  To search for particle events, images from the camera are continuously examined by an online filter that requires a certain number of pixels to reach a certain threshold in the sum of red, green, and blue pixel amplitudes.  The minimum pixel multiplicity has been varied during development but is typically five.

A DECO run begins automatically when the user opens the app and continues until the app is closed.  The amplitude threshold is automatically determined at the beginning of each run by a calibration routine that measures the sensor noise, which varies from device to device and with temperature.  In addition to the filtered data stream of particle candidates, the app saves a single image every five minutes (regardless of whether it passed the filter criteria) as a minimum bias stream for sensor noise and other performance characterization.

Each image contains RGB colors with 8 bits of depth per color.  Complete images are saved in order to characterize background noise fluctuations and resolve the full transition from brightly hit pixels through sub-threshold neighboring pixels down to the thermal noise background.  Users are advised to collect data with the phone plugged in, connected to wifi, and with the camera lens covered in order to minimize ambient light reaching the sensor.

A map of DECO acquisition locations is shown in Figure~\ref{map}.  DECO users have taken data on all seven continents with more than different models of Android devices.


\section{Method of depletion thickness measurement}

At sea level, the cosmic-ray muon flux is dominated by minimum ionizing particles, with a mean energy of $\sim$4~GeV~\cite{PDG}.  The zenith angle ($\theta$) distribution is proportional to $\cos^2 \theta$ and the total flux integrated over energy and solid angle is $\sim$1 particle per minute per cm$^2$~\cite{PDG}.

We approximate the depletion region as uniformly sensitive to ionizing radiation.  Under this assumption, the depletion thickness determines the distribution of the measured track length (projected on the image plane).  Given a depletion thickness ($H$) and track length component in the sensor plane ($L$), the distribution of measured track lengths, in the case of an isotropic particle flux  (for example, dominated by alpha particles from radioactive decay within or near the phone), is given by

\begin{equation}
\frac{dN}{dL} = A \frac{L H^2}{(L^2 + H^2)^{2}}  {\rm~~~(isotropic)}
\end{equation}

where $A$ is a normalization constant proportional to the absolute particle flux, sensor detection efficiency, and livetime.



Alternatively, if the tracks are distributed according to the sea-level cosmic-ray angular distribution ($\cos^2\theta$), a different length distribution is expected:

\begin{equation}
\frac{dN}{dL} = B \frac{L H^4}{(L^2 + H^2)^{3}} {\rm~~~(cosmic~rays)} 
\end{equation}


where B is analogous to A.  See the appendix (Section~\ref{appendix}) for derivations of these distributions.



The two distributions have slightly different shapes that can be distinguished from one another with sufficient statistics.



\section{Event classification and selection}

\begin{figure}[htbp]
\includegraphics[width=0.5\textwidth]{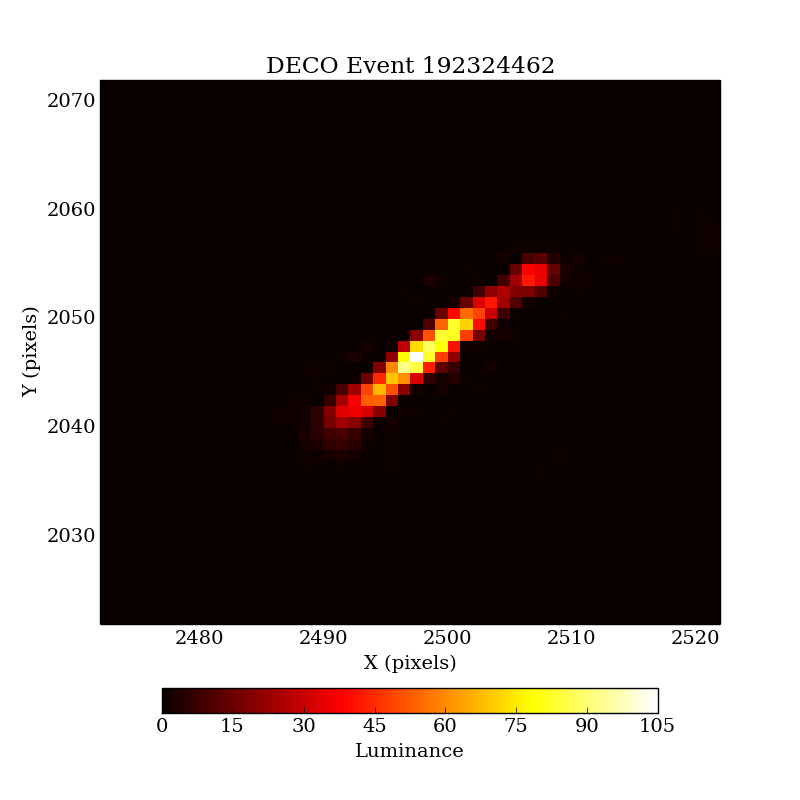}
\caption{Example ``track'' event likely due to a GeV cosmic-ray muon.\label{track}}
\end{figure}

\begin{figure}[htbp]
\includegraphics[width=0.5\textwidth]{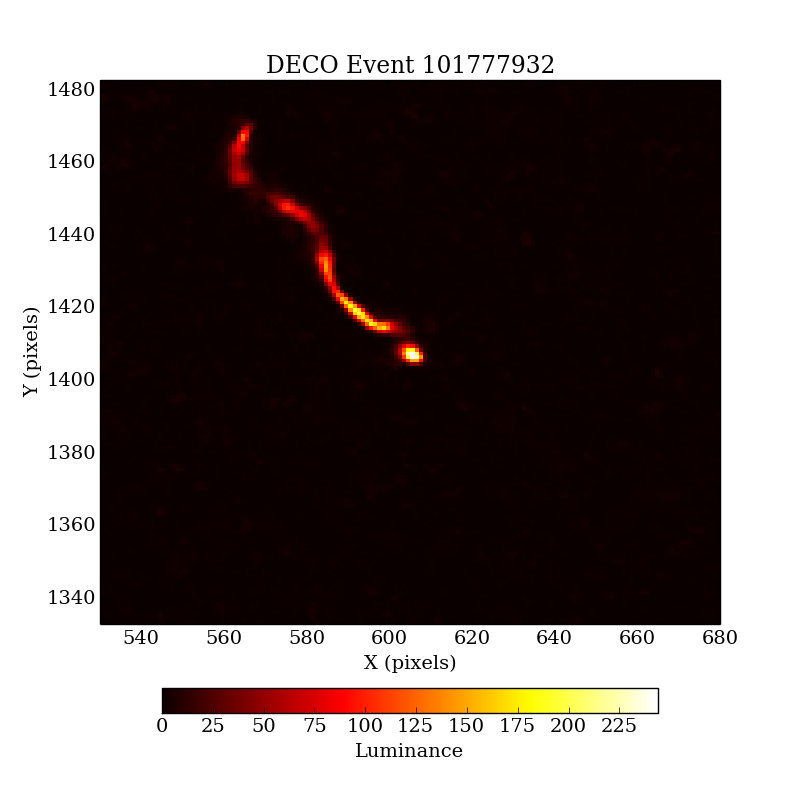}
\caption{Example ``worm'' event likely due to a low-energy (MeV) electron, potentially produced by a Compton scatter from an incident gamma-ray produced by radioactive decay in the phone or surroundings.\label{worm}}
\end{figure}

\begin{figure}[htbp]
\includegraphics[width=0.5\textwidth]{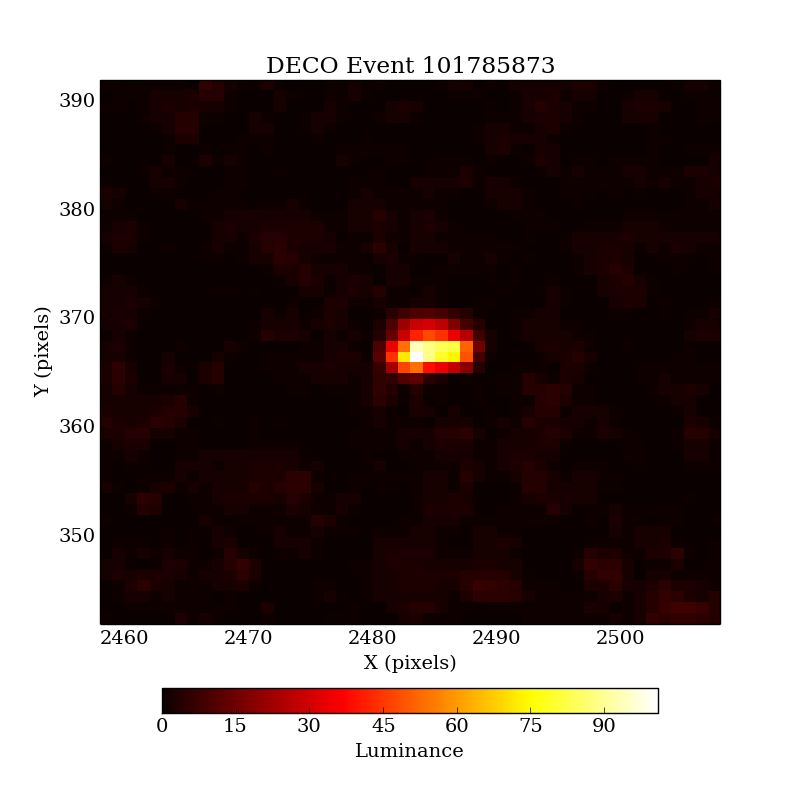}
\caption{Example ``spot'' event.  The spot topology is likely due to a Compton scatter event that produces a low energy (MeV) electron which is quickly absorbed.  \label{spot}}
\end{figure}

Events representative of the three topologies commonly detected by DECO are shown in Figures~\ref{track},~\ref{worm}, and~\ref{spot}.  These event classes detected by DECO in cell phone CMOS image sensors are the same as those identified in astronomical CCDs.  Following the terminology of ~\cite{Groom}, we designate these event classes as ``tracks'', ``worms'', and ``spots''.  Tracks are straight, likely due to high energy (GeV scale) cosmic-ray muons that exhibit little Coulomb scattering.  Worms show significant deflection along the trajectory, curving and/or kinking.  They are most likely due to low-energy (MeV scale) electrons that exhibit multiple Coulomb scattering.  Such electrons are a product of radioactive decay in the materials of the phone itself or in the surroundings.  They could be produced directly in a beta decay event or by a Compton scatter from an incident gamma ray produced in a decay.  Spots could also be produced by short-range alpha particles~\cite{Estrada2011}.

Alpha particles produced by nuclear decay could also deposit detectable tracks.  Due to their short range, they would need to be produced within the phone itself or within a few cm of it.  The range of alpha particles between 1 and 10~MeV in silicon is dozens of microns, comparable to the expected sensor depletion thickness.  Future studies may enable identification of alpha tracks through their ionization energy loss rate ($dE/dx$), which is four times larger than that of muons and electrons due to their twice larger charge, or through their Bragg peak as discussed below.

Some tracks are brighter at one end than the other.  In the case of stopping particles, this could indicate a Bragg peak.  While the probability of cosmic-ray muons stopping within the sensor is negligible due to their long range, alpha particles have a good likelihood of stopping within the depletion region.  It is also possible that the bright spots on one end are due to front-back asymmetry in the response of the depletion region to ionizing radiation~\cite{DAMIC2012}.  If confirmed, this effect could be used to break the bilateral degeneracy in direction reconstruction along the track.

In addition to the depletion thickness measurement presented here using the muon candidate sample, a similar measurement could be performed using the worm events by quantifying their multiple scattering within the depletion region as well as their trajectory into and out of the depletion region slab.

To classify events, we first convert the RGB pixel values to a single gray-scale amplitude (luminance) using the ITU-R 601-2 luma transform~\cite{ITU}.  We next calculate a contour using the ``marching squares'' algorithm~\cite{marching}, implemented in the scikit-image python module~\cite{scikit}, to delimit the pattern of pixels that detected ionization above a particular threshold.  This algorithm interpolates within pixels to determine the best iso-luminance contour through each pixel based on the luminance of it and its neighbors.  The algorithm requires a single parameter which is the luminance value at which the iso-luminance contour should be drawn.  The value found to perform best in containing the hit pixels and not being susceptible to noise was 20.

Several metrics are calculated for each event.  The first is the integrated area within the iso-luminance contour.  To remove background events caused by noise fluctuations, we require a minimum area of 10 pixels.  We also determine the maximum and minimum $x$ and $y$ values spanned by the contour and define the distance between these to be the track length.

Next we calculate the principal moments of the luminance image.  From these, we determine the image eccentricity $\epsilon$.  We find this eccentricity parameter to to be an excellent discriminant for separating the long, straight tracks from worms and spots.  To select a sample rich in muon candidate events, we require $\epsilon > 0.99$.

\begin{figure*}	
\subfiglabelskip=0pt
\begin{center}
\subfigure[][]{
\label{wildfire_isotropic}
\noindent\includegraphics[width = 0.45 \textwidth]{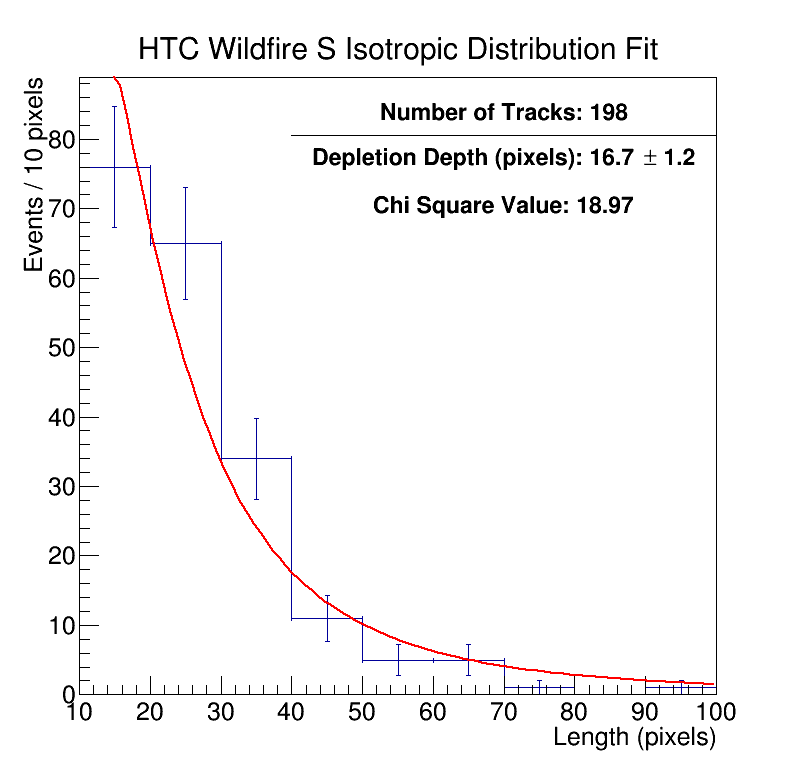}
}
\subfigure[][]{
\label{wildfire_cosmic}
\noindent\includegraphics[width = 0.45 \textwidth]{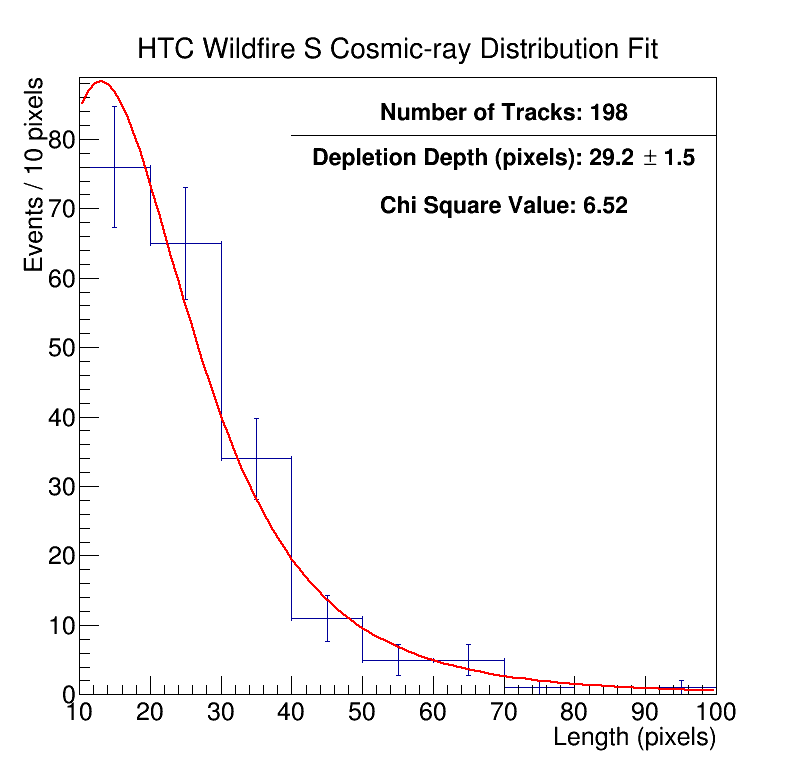}
}
\caption{Measured length distribution of tracks from muon candidates detected by the HTC Wildfire S.  In each panel the distribution is fit with a single-parameter function whose parameter is the sensor depletion thickness in units of the pixel width.  In each case there are $8-1=7$ degrees of freedom.  \subref{wildfire_isotropic} Fit with functional form according to an isotropic particle flux.  The $p$ value for the fit is 0.008.  \subref{wildfire_cosmic}  Fit with functional form according to a sea-level cosmic-ray angular distribution ($\cos^2\theta$).  The $p$ value for the fit is 0.48.
\label{fits}}
\end{center}
\end{figure*}

\section{Results}

     
    In this analysis we focus on 5829 events recorded by the HTC Wildfire S between January 28, 2012 and January 21, 2014.  
    After requiring a minimum area of 10 pixels, 2018 events remain.  We next apply a cut designed to remove worms, noise fluctuations, and multi-site depositions: We require that the marching squares algorithm determines only one closed contour in the entire camera image.  This reduces the sample to 1043 events.  Finally, requiring the eccentricity to be at least 0.99 reduces the number of events to 198.
    
    Using these events we determined the length distribution (in pixels) and fit both the isotropic and cosmic-ray parameterizations to the measured distribution, as shown in Figure~\ref{fits}.  In each case (isotropic and cosmic-ray), a single-parameter fit provides a depletion thickness estimate.  The estimate is 29.2 $\pm$ 1.5 pixels in the cosmic-ray case and 16.7 $\pm$ 1.2 in the isotropic case.  According to~\cite{wildfire}, this camera image sensor is 2400~$\mu$m $\times$ 1800~$\mu$m for its 2592 $\times$ 1944 pixels, corresponding to a 0.9~$\mu$m square pixel size.  This means that the DECO estimated depletion thickness is 26.3 $\pm$ 1.4~$\mu$m (cosmic-ray case) or 15.0 $\pm$ 1.1~$\mu$m (isotropic case).
    
    The $p$ value of the cosmic-ray fit is 0.48 and the $p$ value of the isotropic fit is 0.008.  The present dataset is therefore fit moderately better by a cosmic-ray distribution than an isotropic distribution.  This gives us additional confidence that we are detecting and accurately selecting cosmic-ray muons in the DECO data.  With future datasets it will be possible to more significantly distinguish between a cosmic-ray and isotropic distribution.  This will provide additional confidence in the cosmic-ray origin of the track-shaped signals and will also enable a quantification of a possible isotropic background component produced for example by alpha decays.
    
    Although the event selection is developed to obtain a muon-rich sample and the cosmic-ray fit is preferred over the isotropic fit, there could be an isotropic particle population that contaminates the cosmic-ray muons.  The isotropic result therefore provides an estimate of the systematic uncertainty of the measurement.  Conservatively, the depletion thickness could be within the full range spanned by the two fits: $(13.9, 27.7)$~$\mu$m.

\section{Conclusion}

We have demonstrated a particular capability of a cell phone app and central database, DECO, to detect cosmic-ray muons and other ionizing radiation.  We identified several classes of commonly occurring events consistent with those found in professional astronomical CCDs, and developed algorithms to classify the events.  Using a muon track sample, we used DECO to measure the thickness of the depletion region in camera image sensors of an example cell phone model, the HTC Wildfire S, under the approximation that the depletion region responds uniformly to ionizing radiation.

We note that the track length distribution is different for cosmic rays at sea level than for isotropic particles.  The cosmic-ray fit is preferred over the isotropic fit with moderate significance.  Larger data sets available in the future will more powerfully discriminate between the two cases.


The quality of the directional information of muon tracks in cell phones indicates that additional future measurements are feasible.  Because most modern cell phones have accurate orientation sensors, the absolute direction of each cosmic-ray muon can be determined by fitting the azimuthal angle within the sensor plane and using the track length (with the depletion thickness, which can eventually be measured for each device) to determine the polar angle relative to the sensor plane.  The depletion thickness relative to the pixel width, not the absolute depletion thickness, is necessary for such direction reconstruction and is determined by the method presented here.

We also note that bright spots at one end of many tracks may be due to front-back asymmetry within the active region and may therefore enable breaking the bilateral degeneracy in determining the cosmic-ray incident direction along the track.  It may be possible in future studies to use mobile devices to detect the East-West effect~\cite{Johnson1933, AlvarezCompton1933} in cosmic rays, an effect which historically provided the first indication that cosmic rays are positively charged.  DECO provides a large cosmic-ray monitoring network that could enable correlation studies with other data sets and events such as solar storms.  Although exceedingly difficult~\cite{criticism}, it may also be possible to detect extensive air showers produced by ultra-high-energy cosmic rays through detection with sufficiently many devices in coincidence~\cite{crayfis}.  In any case, DECO is a powerful tool for students and members of the public to carry a particle detector in their pocket and use it to engage with astronomy and particle physics.


\section{Appendix: derivation of parameterizations}

To derive the functional form of the parameterization in the two cases, we consider the dependence on $H$ and $L$ and do not include overall normalization constants.

\subsection{Isotropic case}

In the isotropic case, we have

\begin{equation}
 dN \propto A_\textit{eff}(\theta) d\Omega 
\end{equation}

where $A_\textit{eff}$ is the effective area that the sensor presents to the particle flux.  It is a function of $\theta$ ($A_\textit{eff} = a \cos \theta$, where $a$ is the sensor geometrical area) because the sensor presents a greater cross sectional area to the flux at normal incidence than the flux at oblique incidence.  Furthermore, $d\Omega = \sin \theta d\theta d\phi$.  Therefore,


\begin{equation}
 dN \propto \cos \theta \sin \theta d\theta 
\end{equation}

\begin{equation}
dN \propto d ( \sin^2 \theta) 
\end{equation}

Now, $\sin \theta = L/\sqrt{L^2 + H^2} $, so

\begin{equation}
\sin^2 \theta = L^2 / (L^2 + H^2) 
\end{equation}

\begin{equation}
 d(\sin^2\theta) = \frac{2L H^2 dL}{(L^2+H^2)^2} 
\end{equation}

Therefore,



\begin{equation}
\frac{dN}{dL} = A \frac{L H^2}{(L^2 + H^2)^{2}}  {\rm~~~(isotropic)} 
\end{equation}

\subsection{Cosmic-ray case}
In the cosmic-ray case, we have an additional two factors of $\cos\theta$:

\begin{equation}
 dN \propto \cos^3 \theta \sin \theta d\theta
\end{equation}

\begin{equation}
 dN \propto d ( \cos^4 \theta) 
\end{equation}

And

\begin{equation}
 d(\cos^4\theta) = \frac{-2(L^2+H^2) 2L H^4 dL}{(L^2+H^2)^4} 
\end{equation}

So

\begin{equation}
\frac{dN}{dL} = B \frac{L H^4}{(L^2 + H^2)^{3}} {\rm~~~(cosmic~rays)} 
\end{equation}




\label{appendix}



\section{Acknowledgments}
DECO is supported by the American Physical Society, the Knight Foundation, and the Simon-Strauss Foundation.  We are grateful for valuable conversations with Keith Bechtol, Andy Biewer, Patricia Burchat, Duncan Carlsmith, Alex Drlica-Wagner, Lucy Fortson, Stefan Funk, Mandeep Gill, Giorgio Gratta, David Kirkby, Carsten Rott, David Saltzberg, Ignacio Taboada, and Ian Wisher.  We appreciate beta testing, data collection, and data studies by Colin Adams, Ilhan Bok, Paul Brink, Felipe Campos, Alex Diebold, Mike DuVernois, Laura Gladstone, Jim Haugen, Kyle Jero, and Heather Levy.


\bibliographystyle{elsarticle-num}
\bibliography{references}

\end{document}